# On stress concentration in nanotwinned metals

S. Mohadeseh Taheri-Mousavi[*,&], Huajian Gao[&,$,#,1]

[*]Department of Mechanical Engineering and Department of Materials Science and Engineering, Massachusetts Institute of Technology, 77 Massachusetts Avenue, Cambridge, MA 02139, USA.

[&]School of Engineering, Brown University, Providence, Rhode Island 02912, USA

[$]School of Mechanical and Aerospace Engineering, College of Engineering, Nanyang Technological University, 70 Nanyang Drive, Singapore 639798, Singapore

[#]Institute of High Performance Computing, A*STAR, Singapore 138632, Singapore

In their papers, Lu *et al.*[1] analyzed experimental data from *in-situ* transmission electron microscopy (TEM) and concluded that local stress concentration at grain boundary-twin boundary (GB-TB) intersections in nanotwinned (nt-) metals increases as the TB thicknesses is reduced (below $\lambda \leq 18$ nm). In spite of the elegant experimental study, this conclusion is, however, contradictory to a commonly expected behavior that local shear stress $\tau$ near such an intersection should scale as[2,3]

$$\tau \propto \tau_0 \left(\frac{L}{r}\right)^\alpha \qquad (1)$$

where $\tau_0$ is a reference stress depending on loading and geometrical parameters, $r$ is the distance from the GB-TB intersection, $L$ is the TB spacing, and $\alpha > 0$ is a singularity exponent depending on the detailed crystal orientations and material constants. This scaling law indicates that, as long as there is stress concentration (i.e., $\alpha > 0$), the local shear stress should decrease as the TB spacing is reduced, contrary to the conclusion made by Lu *et al.*[1].

The above conflict has motivated us to re-analyze the raw experimental data by Lu *et al.* We found that, due to the high stress gradient near the GB-TB intersections, the averaging method used by Lu *et al.*[1] for measuring local stress is inadequate for accurately capturing local stress concentration and thus led to the apparent contradiction to the scaling law in Eq. (1).

---

[1] Corresponding author. e-mail: huajian_gao@ntu.edu.sg (H.G.)

To demonstrate this inadequacy, the experimental strain distributions at GB-TB intersections in twin lamellae with twin thicknesses of λ = 1.2 nm and 3.2 nm (see the red arrows and the red curves in Fig. 1a), were extracted from the original paper by Lu et al.[1]. As shown in Fig. 1b, smaller box sizes whose edges are 0.1, 0.2, and 0.6 of the red box edge used by Lu et al. were considered. For any box sizes smaller than the ones used in Lu et al.[1], as the size decreases, , the average shear strain near GB-TB of a larger twin lamella can achieve higher values than that in a smaller twin thickness. Therefore, the average strain is highly dependent on the averaging methodology, and a completely reverse trend can be obtained based on the same raw experimental data.

To reaffirm the scaling behavior in Eq. (1), we have further conducted finite element (FE) simulations on a continuum model containing a spherical nt-grain located at the center of a cube (see Fig. 2a). The edge length of the cube is twice the diameter of the grain. Two samples, each with a spherical nt-grain (see Fig. 2b) consisting of alternate twins and matrices with similar twin thicknesses to those analyzed by Lu et al.[1] (see Fig. 1a), were modeled. The two samples were meshed with the same element size (≈ 0.22 nm). The crystal orientation of the nt-grains is indicated in Fig. 2a. The surrounding crystal has a misorientation angle of 39º, identical to the experimental samples. An anisotropic constitutive law for Cu was considered in the FE simulations (see Table 1). The simulation box is periodic in all three directions and subjected to a unit displacement applied in the XY plane. The direction of the applied displacement forms an inclination angle of 35º with respect to the X axis as in the experiment. The shear stress $\tau_{xy}$ distribution along TB lamellae and the $\tau_{xy}$ contour are shown in Figs. 2b and 2c, respectively, for the two samples. It is seen that the shear stress near the GB-TB intersection of the sample with smaller twin thickness attains lower magnitudes than that with larger twin thickness. As shown in Fig. 1a, the strain distribution at the GB-TB intersection with smaller twin spacing (see the left red arrow) reaches a maximum magnitude of 8% (see the left red curve), which is also smaller than the maximum value of 10% (see the right red arrow and the curve) observed at larger TB spacing. As shown in Fig. 2b, the stress converges to the same magnitude in the middle of the grains, indicating a higher stress concentration at GB-TB intersections in the grain with larger twin spacing. As shown in Fig. 2c, our FE simulations are consistent with the scaling law in Eq. (1) in that the local shear stress near a GB-TB intersection decreases as the twin spacing is reduced.

To analyze the averaging technique used by Lu et al.[1], the stress distribution in three boxes with edge size having different ratios A of the grain size and twin thickness (0.5, 0.25, and 0.1) is presented in Fig 2d. The distribution, the average value, and the cumulative distribution function (CDF) of the shear stress change considerably for three box sizes. This change shows that due to the high stress/strain gradient at GB-TB intersections, the average stress is highly dependent on the adopted box size and therefore may not precisely capture the maximum shear stresses in the sample. Moreover, as shown in Fig. 2e, due to the anisotropy and the shape of the nt-grain, an analysis of stress concentration is highly dependent on the location of the twin lamella. The maximum shear stress at GB-TB intersections in the two samples with a similar twin thickness (see Fig. 2e) has different magnitudes when the twin lamella is moved in the nt-grains.

To further confirm that the shear stress near GB-TB intersections in a sample decreases as the twin thickness is reduced, a continuum model containing an elliptical shaped nt-grain (see Fig. 3a) with a dimension similar to the one in Fig. 3d was constructed. The major and minor axes of the grain, the twin thicknesses, the GB misorientation angle of 35.3º, and the loading direction of 35º with respect to the twin plane were chosen to be identical to the experimental sample. The shear stress along three twin lamellae and shear stress contour are shown in Fig. 3b and 3c, respectively. The shear stress near GB-TB intersection 1, which is near a larger twin lamella (10.75 nm), has a higher magnitude than that near GB-TB intersection 2, which is adjacent to a smaller twin spacing (2.27 nm). This is in fact consistent with the shear strain contour in the experimental sample in Lu et al.[1] (see Fig. 3d) and also the report of the first partial dislocation emission from GB-TB intersection 1 due to higher stress concentration. Therefore, we identify that the magnitude of shear stress concentration measured in the red box cannot be assigned to the twin lamella with a thickness 2.27 nm in the experimental sample (as in Lu et al.[1] paper), as the twin plane under attention has unequal distance to its two adjacent twin planes.

The analysis of the raw experimental data in Lu et al.[1] together with the above simulations and dimensional analysis indicate that measurement of local stress concentration in a nt-material is highly sensitive to the averaging box size due to the high stress gradient near GB-TB intersections as well as to the TB location in the nt-grain. As a consequence, the authors could not provide a convincing case by ruling out all other influential parameters and thus the averaging technique adopted in Lu et al.[1] cannot reliably capture the local stress concentration. Moreover, a measured stress intensity can be assigned to a chosen GB-TB intersection only if the two adjacent twin lamellae of the selected twin plane have similar thicknesses. Our FE simulations, which is consistent with the widely known scaling law in Eq. (1), show that the local stress intensity near a GB-TB intersection must decrease as the TB spacing is reduced. The above discussions reaffirm the applicability of continuum mechanics in analyzing stress concentration in nanostructured materials and also highlight an urgent need to develop more reliable methods to accurately extract the local stress concentration in microstructures with high stress gradient, as encountered not only in the *in-situ* TEM analysis in Lu et al.[1], but also in other popular data analysis techniques such as digital image correlation (DIC) for scanning electron microscopy (SEM) and optical images.

All data generated and analyzed during this study are included in this published article.

## Acknowledgments

The authors acknowledge financial support from Swiss National Science Foundation through Grant P2ELP2_162144 (to S.M.T.M.) and National Science Foundation through grant DMR-1709318 (to H.J.G.). The reported simulations were performed on resources provided by the Extreme Science and Engineering Discovery Environment (XSEDE) through Grant MS090046. Helpful discussions with B. Ni is gratefully acknowledged. We also acknowledge a helpful discussion with Prof. Lei Lu from Shenyang National Laboratory for Materials Science, Institute of Metal Research, Chinese Academy of Sciences.

**Author contributions**

H.J.G. conceived and designed the project. S.M.T.M. performed the finite element simulations. S.M.T.M., H.J.G analyzed the results and co-wrote the manuscript.

**Competing interests**

The authors declare no competing interests.

Table 1. Anisotropic elastic constants for Cu[4]

| Materials | Cu |
|---|---|
| $C_{11}$ (GPa) | 168.4 |
| $C_{44}$ (GPa) | 75.4 |
| $C_{12}$ (GPa) | 121.4 |

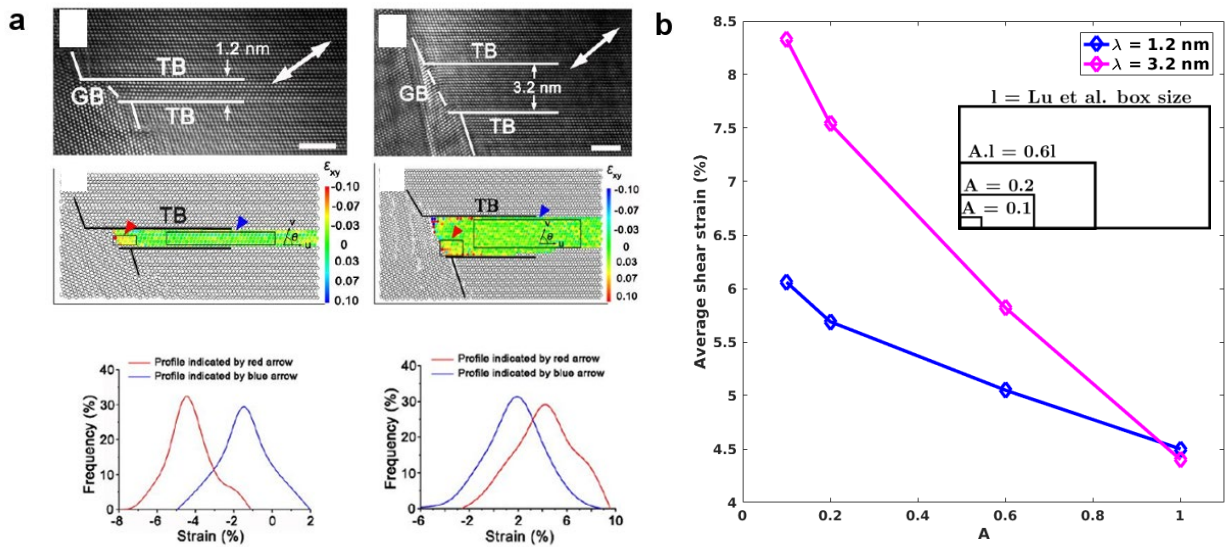

**Fig 1. Average shear strain at grain boundary-twin boundary (GB-TB) intersections of raw experimental data of Lu *et al*.[1] for different box sizes.** a) Raw experimental data from Lu *et al*.[1] showing shear strain contour and distribution in twin lamellae with twin thicknesses of λ = 1.2 nm and 3.2 nm. b) Average shear strain at GB-TB intersections in twin lamella with twin thicknesses of λ = 1.2 nm and 3.2 nm. The box for averaging has an edge which is 0.1, 0.2, and 0.6 of the box edge in Lu *et al*. (see the red arrows and the red curves in Fig. 1a). The average strain achieves higher magnitude near the GB-TB intersection of a large twin thickness when the box size reduces.

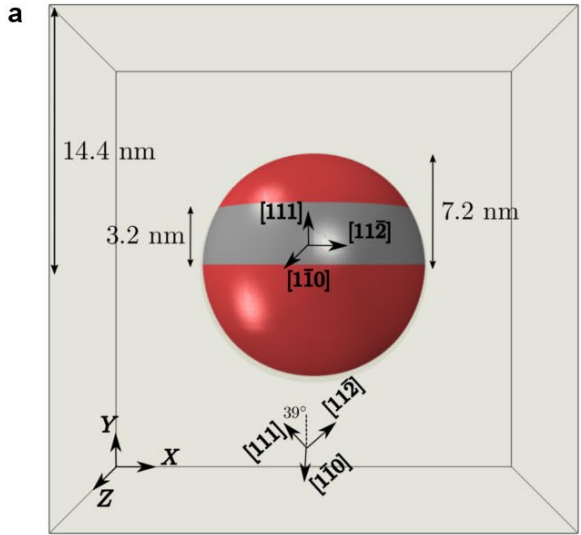
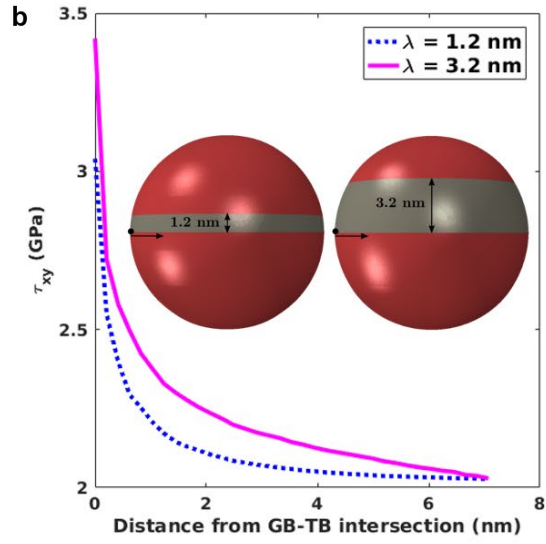
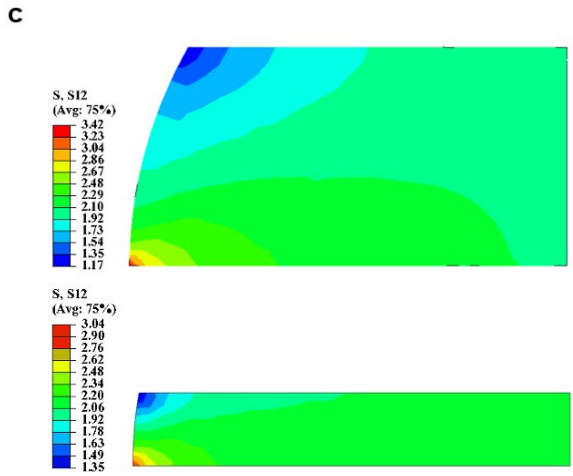
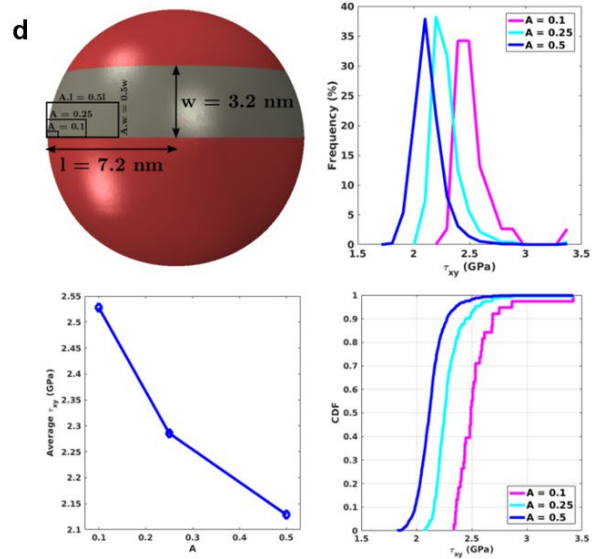
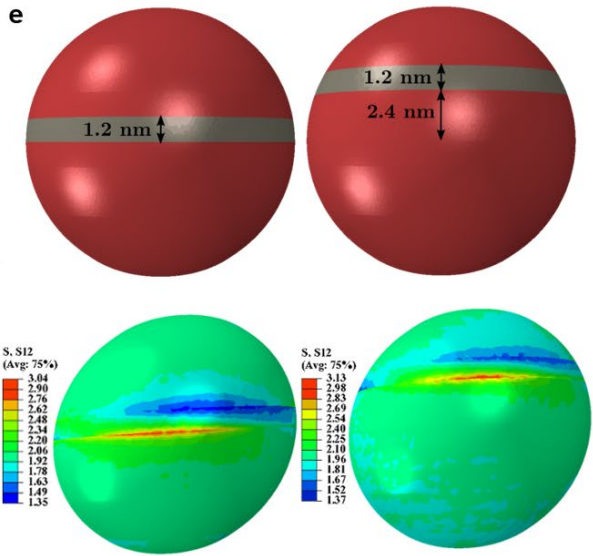

**Fig 2. Finite element (FE) simulations showing high sensitivity of local shear stress concentration on twin thickness, the averaging box size, and the location of the TB in a nanotwinned (nt-) grain.** a) FE model of a spherical nt-grain containing alternate twin and matrices located at a center of a cube. b) Shear stress distribution along twin lamellae of two nt-grains with twin thickness of $\lambda$ = 1.2 nm and 3.2 nm. c) Shear stress contour in the two simulated twin lamellae. d) Shear stress distribution from three boxes with edge size having different ratios A of the grain size and twin thickness (0.5, 0.25, and 0.1). The lower left curve shows average stress in the three boxes. The lower right figure presents cumulative density function (CDF) of shear stress, which exhibits considerably different distribution by changing the size of the averaging box. e) Shear stress contour in two nt-grains with a similar twin thickness, and different locations of the twin lamella. Due to the anisotropy and shape of the grain, the maximum stress is highly dependent on the location of the twin lamella.

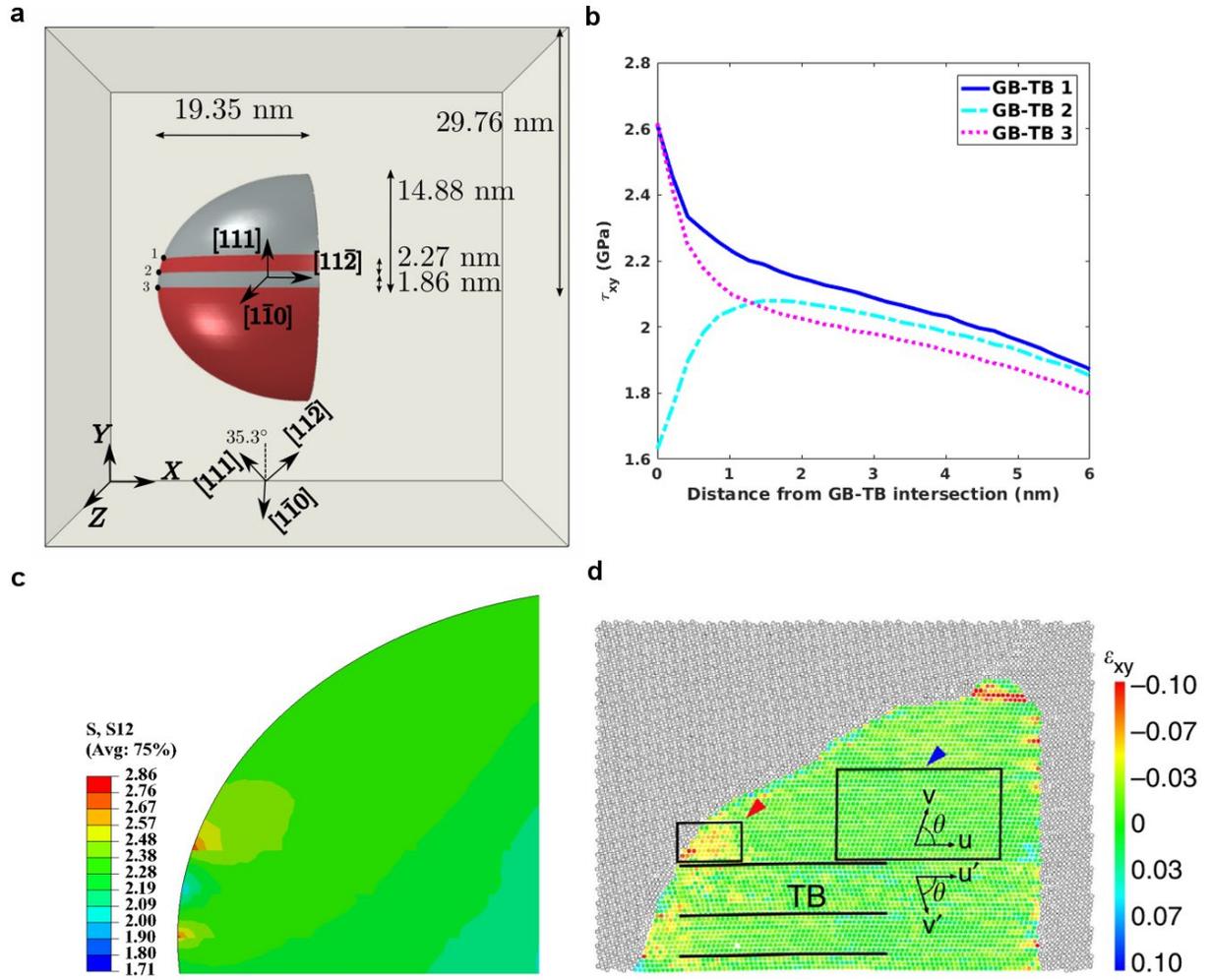

**Fig 3. FE simulations showing higher shear stress concentration at GB-TB intersection of a larger twin thickness.** a) FE model of a half elliptical shaped nt-grain containing alternate twins and matrices located at a center of a cube. b) Shear stress distribution along twin planes from GB-TB intersections 1, 2, and 3 indicated in (a). c) Shear stress contour in nt-grain showing higher stress distribution near GB-TB intersection 1, which is adjacent to a larger twin lamella. d) Shear stress contour from *Lu et al.*[1] confirming higher stress magnitude along GB-TB intersection 1.